\documentclass[prb,twocolumn,showpacs]{revtex4}
\usepackage{amsmath}
\usepackage{dcolumn}
\usepackage{graphicx}

\begin{document}

\title[Short title for running header]{Z$_{2}$ spin liquid phase on the kagome lattice: a new saddle point}
\author{Tao Li}
\affiliation{Department of Physics, Renmin University of China,
Beijing 100872, P.R.China}
\date{\today}

\begin{abstract}
We have performed large scale variational search for the best RVB ansatz for the spin-$\frac{1}{2}$ kagome 
antiferromagnet with both nearest-neighboring(NN) and next-nearest-neighboring(NNN) exchanges assuming 
only translational symmetry. We find the best RVB state is always fully symmetric and has an mean field ansatz 
that is gauge equivalent to a previous proposed Z$_{2}$ spin liquid ansatz. The Z$_{2}$ state is found to be slightly 
more stable than the extensively studied U(1) 
gapless Dirac spin liquid state in both the $J_{2}\neq0$ and the $J_{2}=0$ case and to possess a small spinon gap 
for $J_{2}<0.2$. The breaking of the U(1) gauge symmetry in the Z$_{2}$ state is found to increase with $J_{2}$ and to be
quite substantial around $J_{2}=0.15$. However, we find the Z$_{2}$ state is always very close to the gapless U(1) 
Dirac spin liquid state, although they have very different RVB parameters. We argue the kagome antiferromagnet 
should be better understood as a near critical system, rather than a system deep inside a gapped spin liquid phase 
with well established Z$_{2}$ topological order.
\end{abstract}


\maketitle
The search for spin liquid state in frustrated quantum antiferromagnets has attracted a lot of attention in the strongly 
correlated electron system community. The spin-$\frac{1}{2}$ kagome antiferromagnet 
is a particularly promising system for this purpose because of the strong geometrical frustration and the low coordinate 
number of the lattice. Early studies on small clusters find the ground state of the kagome antiferromagnet with NN 
exchange has no signature of any symmetry breaking and is very likely a quantum spin liquid\cite{Elser,Chalker,Leung,Young,Lecheminant,Sindzingre,Nakano,Lauchli}.
However, the exact nature of such a spin liquid phase is still under intense debate\cite{Series,Vidal}. Especially, it is found that 
the gap for triplet excitation above the ground state is very small, although the spin correlation length is only 
of the order of the lattice constant. More curiously, it is find that the spin singlet channel of the system may host 
an exponentially large number of low energy excitations below the exceptional small spin gap\cite{Singlet,Mila,Auerbach,Poilblanc}. 

The quest for the true nature of the ground state of the kagome antiferromagnet becomes even more urgent after 
its material realization in the 
$\mathrm{ZnCu}_{3}\mathrm{(OH)}_{6}\mathrm{Cl}_{2}$ systems\cite{Helton,Mendels,Han}. Recently, extensive DMRG simulations has been 
performed on the kagome antiferromagnets both with 
only NN exchange and that with more extended exchange couplings\cite{Jiang1,Yan,Depenbrock,Jiang2,Sheng,Kolley,Gong}. The DMRG results seem to indicate that the kagome 
antiferromagnet with NN exchange has a small but finite gap in both spin triplet and spin singlet channel, 
which is in support of a 
gapped Z$_{2}$ spin liquid scenario. However, it is found that the $\ln 2$ topological entanglement entropy expected for 
a gapped Z$_{2}$ spin liquid can only be faithfully demonstrated when one introduce a finite NNN exchange coupling\cite{Jiang2}. 
The abundance of low energy singlet excitations in the system\cite{Singlet} also raises the suspicion that the DMRG results may 
not represent the intrinsic property of the system in the thermodynamic limit, since open boundary 
condition is adopted in all DMRG simulations and symmetry breaking perturbations in the singlet channel 
can be super relevant.

Another scenario on the spin liquid ground state of the kagome antiferromagnets is provided by variational studies
based on the RVB theory. 
It was found previously that the best RVB state for the ground state of the kagome antiferromagnet with NN exchange 
coupling is a U(1) spin liquid state with a Dirac-type spinon dispersion\cite{Hastings,Ran}. Extensive studies on this state find it is
actually a very robust saddle point provided that the symmetry of the system is unbroken\cite{Lu,Iqbal1,Iqbal2}. 
Attempts to break the U(1) gauge symmetry(and to introduce spin gap) always result in increase of energy 
in the thermodynamic limit. More recently, study on the $J_{1}-J_{2}$ Heisenberg model on the kagome lattice has been 
performed and it is found that the U(1) Dirac spin liquid state remains the most stable phase even when $J_{2}$ is 
quite large\cite{Iqbal3}. This is strange since strong evidence for Z$_{2}$ topological order has already been reported 
in DMRG simulation in this case. The U(1) Dirac spin liquid state is thus argued to form a stable phase in the 
phase diagram around the NN Heisenberg model point, rather than a single quantum critical point.

In this work, we reinvestigate the possible spin liquid ground state of the kagome antiferromagnet with the RVB theory. 
To reduce bias in our study, we only assume translational symmetry for the RVB state at the beginning. We have performed 
large scale variational search for the best RVB ground state for the $J_{1}-J_{2}$ model on the kagome lattice 
with up to the second neighbor RVB order parameters. We find the best RVB state is a fully symmetric spin liquid 
with a mean field ansatz that is gauge equivalent to a Z$_{2}$ ansatz proposed previously from the analysis of 
projective symmetry group(PSG) on the kagome lattice\cite{Lu}. 
Different from a previous study on the same state, we find a finite spin gap and Z$_{2}$
gauge structure can be indeed be stabilized in both the $J_{2}\neq0$ and the $J_{2}=0$ case, although the energy 
advantage over the U(1) gapless state is very small. We find the extent of U(1) gauge symmetry 
breaking increases with $J_{2}$. However, the spin gap is found to follow the opposite trend and vanishes around 
$J_{2}=0.2$. The size of the spin gap is found to be very small.

The model we study in this paper is given by
\begin{equation}
H=J_{1}\sum_{<i,j>}\mathrm{S}_{i}\cdot\mathrm{S}_{j}+J_{2}\sum_{<<i,j>>}\mathrm{S}_{i}\cdot\mathrm{S}_{j}
\end{equation}
The motivation to introduce $J_{2}$ is to perturb away the system from the U(1) Dirac spin liquid phase and to stabilize
the possible Z$_{2}$ spin liquid phase\cite{Jiang2,Sheng,Kolley,Gong}. 
Here $<i,j>$ denotes NN pair of sites, $<<i,j>>$ denotes NNN pair of sites. 
To describe the possible Z$_{2}$ spin liquid phase on the kagome lattice, we rewrite the spin operators in terms of 
the slave particle operators $\mathrm{S}=\frac{1}{2}\sum_{\alpha,\beta}f^{\dagger}_{\alpha}\sigma_{\alpha,\beta}f_{\beta}$ 
and introduce the following mean field ansatz for the slave Fermion $f_{\alpha}$
\begin{equation}
H_{MF}=\sum_{i,j}\psi_{i}^{\dagger}U_{i,j}\psi_{j}
\end{equation}
in which $\psi_{i}=(f_{i,\uparrow},f^{\dagger}_{i,\downarrow})$ is the Nambu spinor constructed from the slave Fermion. 
$U_{i,j}=\left(\begin{array}{cc}\chi_{i,j} & \Delta^{*}_{i,j} \\\Delta_{i,j} & -\chi^{*}_{i,j}\\ \end{array}\right)$ 
is a $2\times2$ matrix with its matrix element $\chi_{i,j}$ and $\Delta_{i,j}$ representing the hopping
and pairing type RVB order parameters for the ansatz. The RVB state is constructed by Gutzwiller projection of the 
mean field ground state and is given by
\begin{equation}
|\mathrm{RVB}\rangle=\mathrm{P_{G}}|\{\chi_{i,j},\Delta_{i,j}\}\rangle, 
\end{equation}
here $|\{\chi_{i,j},\Delta_{i,j}\}\rangle$ denotes the mean field ground state of $H_{MF}$ and is a generalized 
BCS-type state. $\mathrm{P_{G}}$ denotes the Gutzwiller projection to remove the doubly occupied configuration in the
mean field ground state. The resultant RVB state can be simulated efficiently with the variational Monte Carlo method.

In our RVB state we will keep RVB order parameters $U_{i,j}$ up to the second neighboring bonds.
To represent a fully symmetric spin liquid state, these RVB order parameters should be invariant 
under the symmetry operations of the Hamiltonian up to some SU(2) gauge transformations of the form 
$U_{i,j}\longrightarrow G_{i}U_{i,j}G^{\dagger}_{j}$. Here $G_{i}$ is a SU(2) matrix acting on the Nambu spinor 
$\psi_{i}$. The symmetric RVB state so constructed can be 
classified into distinct PSGs according to the gauge transformations $\{G_{i}\}$ needed to restore the form of the 
RVB order parameters after the symmetry operations. Here we will only assume the 
translational symmetry at the beginning. According to the scheme of the PSG classification, there are only two ways 
to realize the translational symmetry for a Z$_{2}$ spin liquid\cite{Wen}. In the first class, the mean field ansatz should be
manifestly translational invariant. In the second class, the unit cell of the mean field ansatz should be doubled 
and a Z$_{2}$ gauge transformation is needed to restore the translated ansatz. In this study we will concentrated on the 
second class since it is energetically much more favorable than the first class. It can also be shown that 
only the second kind of RVB state can be connected continuously to the U(1) Dirac spin liquid phase. 

In our study, we will adopt a $L\times L \times 3$ cluster as illustrated in Fig.1. The doubled unit cell is illustrated 
by the region within the pink parallelogram. According to the PSG of the state, the RVB order parameters on the dashed 
bonds should be multiplied by an additional minus sign as compared to the RVB order parameters on the solid bonds
translated from them. The on-site RVB order parameters should be manifestly translational invariant. As a result, 
we have 3 independent on-site RVB order parameters, 6 independent RVB order parameters on the NN bonds and 6 
independent RVB order parameters on the NNN bonds. Using the remaining gauge degree of freedom allowed by the 
translational symmetry, we can transform the RVB order parameters on the gray sites and the RVB order parameters 
on the two NN bonds from the gray sites into the form of $\alpha\tau_{3}$. Finally, we can transform the pairing 
order parameter on a third NN bonds into real.
Taking all these considerations into account, we are left with 48 independent real variational parameters.

We have performed large scale variational optimization on the above RVB ansatz for the $J_{1}-J_{2}$ model on the kagome
lattice. We have adopted the Stochastic Reconfiguration method in our variational search\cite{SR}.
The optimized RVB ansatz is illustrated in Fig.1. More specifically, the on-site RVB order parameter are given by
\begin{eqnarray}
U_{i,i}=\left\{\begin{aligned}
      \mu\tau_{3} & &   \mathrm{gray\ sites} \\
      \mu\vec{n}_{\phi_{1}}\cdot\vec{\tau} & &  \mathrm{navy\ sites}
      \end{aligned}
      \right.
\end{eqnarray}
in which $\vec{\tau}=(\tau_{1},\tau_{2},\tau_{3})$ are the Pauli matrixes,
 $\vec{n}_{\phi}=(\sin(\phi),0,\cos(\phi))$ is a vector of unit length in the $\tau_{1}-\tau_{3}$ plane.
 $\mu$ is a real number.
The RVB order parameters on the nearest-neighboring(NN) bonds are given by
\begin{eqnarray}
U_{i,j}=-s_{i,j}\left\{\begin{aligned}
       \tau_{3} & &   \mathrm{black\ bonds} \\
       \vec{n}_{\phi_{2}}\cdot\vec{\tau} & &  \mathrm{red\ bonds}
      \end{aligned}
      \right.
\end{eqnarray}
Here $s_{i,j}=\pm 1$ on the solid(dashed) NN bonds. 
The RVB order parameter on the next-nearest-neighboring(NNN) bonds are given by
\begin{eqnarray}
U_{i,j}=-\nu_{i,j}\left\{\begin{aligned}
      \eta\vec{n}_{\phi_{3}}\cdot\vec{\tau} & &   \mathrm{blue\ bonds} \\
      \eta\vec{n}_{\phi_{4}}\cdot\vec{\tau} & &   \mathrm{green\ bonds} 
      \end{aligned}
      \right.
\end{eqnarray}
Here $\eta$ is a real number,  $\nu_{i,j}=\pm 1$ on the solid(dashed) NNN bonds. 

\begin{figure}
\includegraphics[width=9cm,angle=0]{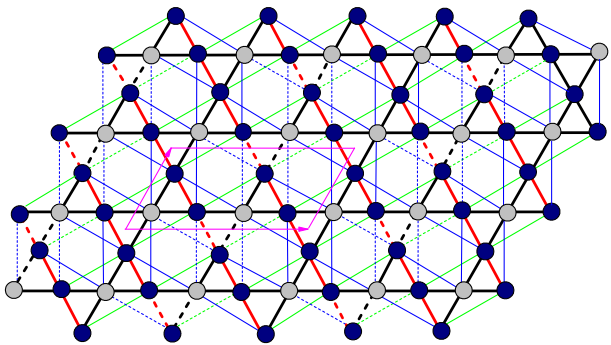}
\caption{The mean field ansatz of the Z$_{2}$ spin liquid state studied in this paper. The region within the pink 
parallelogram is the unit cell of the mean field Hamiltonian. The dots(bonds) in the same color have the same value 
for the RVB order parameter $U_{i,j}$, except for an additional minus sign on the dashed bonds.} \label{fig1}
\end{figure}

The above ansatz is manifestly time reversal symmetric but may in general break the point group symmetry 
of the kagome lattice. However, we find at the optimum of the variational energy, the following
relations always hold: $\phi_{1}=2\phi_{2}$,$\phi_{4}=\phi_{3}+\phi_{2}$. In such a situation, 
we can actually prove that the ansatz describes a fully symmetric spin liquid state.
In fact, we can show that the above RVB ansatz is gauge equivalent to the Z$_{2}$ ansatz proposed in Ref.[1] 
by the following gauge transformation
\begin{eqnarray}
G_{i}=\left\{\begin{aligned}
\exp(-i\frac{\phi_{2}\tau_{2}}{2}) & & i\in   \mathrm{gray\ sites}\\
\exp(i\frac{\phi_{2}\tau_{2}}{2}) & & i\in  \mathrm{navy\ sites}
     \end{aligned}
     \right.
\end{eqnarray}
The gauge transformed ansatz takes the form of
\begin{eqnarray}
U_{i,j}=\left\{\begin{aligned}
\mu\vec{n}_{\phi_{2}} \cdot \vec{\tau} & & \mathrm{on-site} \\
                -s_{i,j}\tau_{3}& & \mathrm{NN\ bonds}\\  
                -\nu_{i,j}\eta\vec{n}_{\phi_{3}} \cdot \vec{\tau}& & \mathrm{NNN\ bonds}
\end{aligned}  
\right.
\end{eqnarray}

When both $\phi_{2}$ and $\eta$ are set to be zero, the above ansatz is reduced to the U(1) Dirac spin liquid ansatz.
To break the U(1) gauge symmetry, we can introduce a nonzero 
pairing term either on a site or on a NNN bond. More specifically, if there exists two closed loops 
starting from the same point, say, site $i$, on which the successive product of the RVB order parameters 
$P_{i}=U_{i,j}U_{j,k}...U_{l,i}$ do not commute, the U(1) gauge symmetry is broken. To determine the gauge structure of
the above ansatz, we can define the following two closed loops. The first one is a length-zero loop consisting of
the site $i$ only. The product of RVB order parameters on this loop is just the on-site RVB order parameter 
$P_{i}^{1}=U_{i,i}=\mu\vec{n}_{\phi_{2}}\cdot\vec{\tau}$. The second closed loop is the elementary triangle from site $i$. 
The product of RVB order parameters on this loop is given by $P_{i}^{2}=U_{i,j}U_{j,k}U_{k,i}=-\tau_{3}$. Here $i,j,k$
are the three vertices of the elementary triangle. Thus, the U(1)
gauge symmetry is broken when $\phi_{2}\neq 0$ or $\pi$\cite{2}. The deviation of $\phi_{2}$ from $0$ or $\pi$ 
provides us a measure of the extent of the U(1) gauge symmetry breaking.
For the above ansatz, the U(1) gauge symmetry is broken when $\phi_{2}\neq 0,\pi$ even in the absence of NNN 
RVB order parameters. However, our numerical optimization shows that the NNN RVB order parameter is crucial to 
stabilize a Z$_{2}$ gauge structure. More specifically, if we set $\eta=0$, then in 
the optimized ansatz we always have $\phi_{2}=0$. As we will see below, the optimized value of both $\phi_{2}$
and $\eta$ increase with $J_{2}$. This is consistent with previous DMRG studies which find that the 
convergence of the topological entanglement entropy becomes better for larger $J_{2}$.

\begin{figure}
\includegraphics[width=9cm,angle=0]{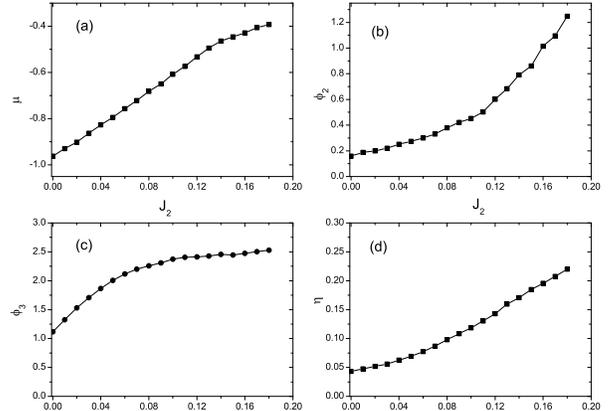}
\caption{The optimized variational parameters as a function of $J_{2}$. The system size is $L=12$ in the calculation}.
\end{figure}

Now we present our results of the variational calculation on the $J_{1}-J_{2}$ model with the above RVB ansatz. 
Our calculation is done on a $L\times L\times 3$ cluster with $L$ ranging from 12 to 24. We use periodic-
anti-periodic boundary condition in all calculations. Previous studies on the same ansatz find that the 
U(1) Dirac spin liquid is the best variational state even for quite 
large $J_{2}$ on sufficiently large system. What we find here is different. We find the U(1) Dirac spin liquid is not 
the most stable state for both zero and non-zero $J_{2}$ on the largest system we have attempted, which is much larger 
than that is used in previous studies. 
In the fully symmetric state, there are four variational parameters, namely $\mu$,
$\phi_{2}$, $\phi_{3}$ and $\eta$ to be determined by optimization of the variational energy. 
In Fig.2 we plot the optimized 
values of the variational parameters as functions of $J_{2}$($J_{1}=1$). 
From the results, we see the extent of U(1) gauge symmetry breaking(which is dictated by the angle 
$\phi_{2}$) increases monotonically with $J_{2}$. At $J_{2}=0.15$, the breaking of the U(1)
gauge symmetry is very significant.

To show that the Z$_{2}$ saddle point we find is indeed more stable than the U(1) saddle point found in previous studies, 
we plot the interpolation of the variational energy between these two saddle points. 
The U(1) saddle point is reached when we set $\phi_{2}=0$ and $\phi_{3}=0$.
The value of $\eta$ is in general nonzero for nonzero $J_{2}$ and should be optimized. We define an interpolation between 
the optimized U(1) ansatz and the optimized Z$_{2}$ saddle point by mixing their variational parameters 
in the following way: $x_{i}=\alpha x_{i}^{(0)}+(1-\alpha)x_{i}^{(1)}$. 
Here $\alpha$ is an interpolation parameter, $x_{i}^{(0)}$ and
$x_{i}^{(1)}$ denote the set of optimized variational parameters($\mu,\phi_{2},\phi_{3},\eta$) for the U(1) and 
Z$_{2}$ saddle point. The variational energy as a function of the interpolation parameter $\alpha$ at $J_{2}=0.15$ 
is shown in Fig.3. The Z$_{2}$ saddle point is obviously more 
robust than the U(1) saddle point. However, the energy difference between the two saddle points is extremely 
small(or the order of $10^{-4}J_{1}$ per site). This is very unusual since the difference in the gauge structure 
of the two saddle points is so significant.

\begin{figure}
\includegraphics[width=8cm,angle=0]{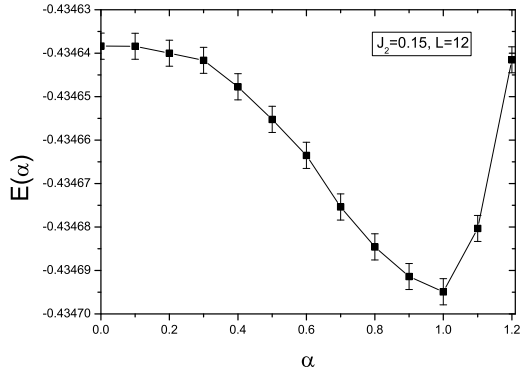}
\caption{The interpolation of the variational energy between the U(1) saddle point and the Z$_{2}$ saddle point. 
The calculation is done for $J_{2}=0.15$ on a $L=12$ system.}
\end{figure}
\begin{figure}
\includegraphics[width=8cm,angle=0]{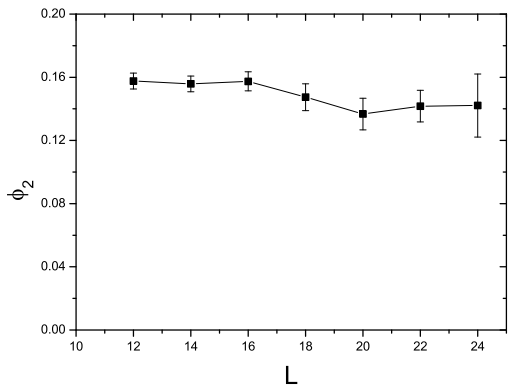}
\caption{The optimized value of $\phi_{2}$ at $J_{2}=0$ for different system sizes.}
\end{figure}

At the same time, we find the Z$_{2}$ phase angle $\phi_{2}$ 
remains nonzero even for $J_{2}=0$(note the optimized value for $\eta$ is significantly
larger than previous report, which is $0.0186(2)$\cite{Iqbal1}). To check if this is a finite size effect, 
we have performed the variational optimization on larger systems. Fig.4 shows the size dependence of 
$\phi_{2}$ up to $L=24$, which is the largest size that we can deal with. We find although there is a tendency 
in $\phi_{2}$ to decrease with increasing $L$, the slope is very small and it is hard conclude if it vanish on 
any finite system that we can treat. We note the largest $L$  we have treated is already significantly
larger than the circumference of the cylindrical clusters used in DMRG calculations. Thus it is possible that the 
weak signature of Z$_{2}$ spin liquid claimed by DMRG simulations at $J_{2}=0$ is still a finite size effect. 
The true ground state at $J_{2}=0$ in the thermodynamic limit may still be very close to a gapless U(1) spin liquid, 
or a quantum critical point\cite{Tao}.

Now let's investigate the evolution of the spin gap with $J_{2}$. Usually, the spin gap opens when 
the mean field ansatz breaks the U(1) gauge symmetry.  We thus should expect the spin gap to increase with the 
extent of U(1) gauge symmetry breaking. However, we find this is not true for our state. 
In principle, it is impossible to extract the spin gap directly from the 
variational calculation on the ground state.
The spin correlation length in the ground state is also not a good indicator for the spin excitation gap 
for the kagome antiferromagnet. Here we will use the mean field result for the spinon gap as an indicator of the
spin excitation gap. The evolution of the mean field spinon gap with $J_{2}$ is shown in Fig.5
(note the magnitude of the NN RVB order parameter has been set to be one). To reduce finite size effect,
we have calculated the mean field dispersion in the thermodynamic limit with the optimized RVB parameters 
obtained on the $L=12$ cluster. We find the spinon gap actually decreases with increasing $J_{2}$ and approaches
zero around $J_{2}=0.2$, a signature which may imply the proximity of the system to a magnetic ordered state 
at large $J_{2}$\cite{Gong}.
\begin{figure}
\includegraphics[width=8cm,angle=0]{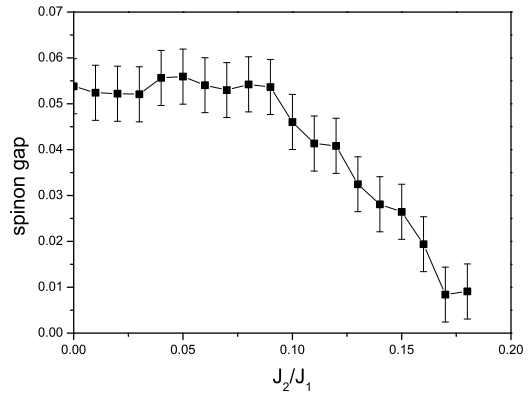}
\caption{The mean field spinon gap in the thermodynamic limit calculated with the optimized RVB parameters obtained on 
the $L=12$ cluster.}
\end{figure}

Finally, let us return to the problem of the exceptional insensitivity of the variational energy in the RVB parameters 
as we find in Fig.3. Similar behavior has also been reported in previous variational studies of the kagome antiferromagnet.
A possible origin for such a strange behavior is that the saddle points with distinct RVB parameters may 
correspond to similar spin liquid state. This implies that some RVB parameters may be quasi-redundant.
To check if this is the case, we have calculated the overlap between the U(1) 
spin liquid state and the Z2 spin liquid state at $J_{2}=0.15$. To our surprise, we find the overlap is larger than 0.99 
on the $L=12$ cluster(the overlap at $J_{2}$ is even larger). To find out the origin for such a quasi-redundancy, 
we plot in Fig.6 the density of state of spinon excitation for the two ansatzs at $J_{2}=0.15$. Except for the small gap 
feature around $E=0.03$ for the Z$_{2}$ ansatz, the density of state of both ansatzs are almost 
identical below $E=1$. The density of state for both states become drastically different only when $E>1$. 
This implies that the main difference between the two states is in their high energy behavior, which for some reason
depend only very weakly on the structure of the ground state. This is possible for a multi-band system, in which
totally different high energy excitation spectrum can be realized on the same ground state. 
It is interesting to see if similar behavior also occurs for RVB states on other on-primitive Bravais lattices.

\begin{figure}
\includegraphics[width=8cm,angle=0]{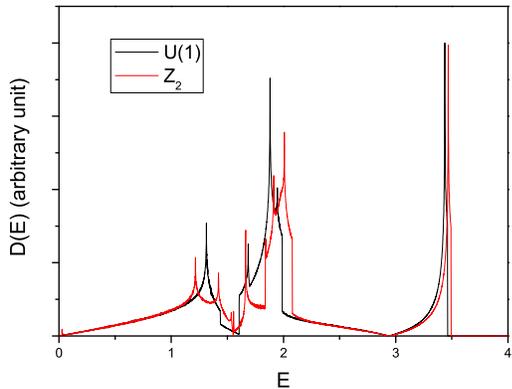}
\caption{The density of state of the U(1) and Z$_{2}$ ansatz at $J_{2}=0.15$.
}
\end{figure}

In summary, we have performed large scale search for Z$_{2}$ spin liquid state for the spin-$\frac{1}{2}$ kagome 
antiferromagnet with NN and NNN exchanges assuming only translational invariance. We find the best RVB state is 
always fully symmetric and is gauge equivalent to a Z$_{2}$ ansatz proposed previously from symmetry analysis. The
Z$_{2}$ state is found to be slightly more stable than the extensively studied U(1) gapless Dirac spin liquid state in both 
the $J_{2}\neq0$ and the $J_{2}=0$ case and to possess a small spinon gap for $J_{2}<0.2$.
We find while the extent of U(1) gauge symmetry breaking in the Z$_{2}$ state increases with 
$J_{2}$, the spinon gap follows the opposite trend.
These results are qualitatively consistent with the findings of the DMRG simulations on the $J_{1}-J_{2}$
model on the kagome lattice. However, we note the Z$_{2}$ state we find is always very close to the gapless U(1) 
Dirac spin liquid state, although they have very different RVB parameters. Thus we think the kagome antiferromagnet 
should be better understood as a near critical system, rather than a system deep inside a gapped spin liquid phase 
with well established Z$_{2}$ topological order.

This work is supported by NSFC Grant  No. 11034012 and the Research Funds of Renmin University of China. 
The author acknowledge the discussion with Tomonori Shirakawa, Seiji Yunoki, Yuan-Ming Lu, Normand Bruce and Fa Wang 
in different stages of this work.

\end{document}